\documentstyle[aaspp4,11pt]{article}

\slugcomment{accepted for publication in ApJ}

\begin{document}

\title
{A Local One-Zone Model of MHD Turbulence in Dwarf Nova Disks}

\author
{Takayoshi Sano\altaffilmark{1,2} and 
James M. Stone\altaffilmark{1,2}}

\altaffiltext{1}
{Department of Astronomy, University of Maryland, 
College Park, MD 20742-2421}

\altaffiltext{2}
{Department of Applied Mathematics and Theoretical Physics, 
Centre for Mathematical Sciences, University of Cambridge, 
Wilberforce Road, Cambridge CB3 0WA, UK; 
T.Sano@damtp.cam.ac.uk, J.M.Stone@damtp.cam.ac.uk}

\begin{abstract}
The evolution of the magnetorotational instability (MRI) during the
 transition from outburst to quiescence in a dwarf nova disk
 is investigated using three-dimensional MHD simulations.
The shearing box approximation is adopted for the analysis, so that the
 efficiency of angular momentum transport is studied in a small local
 patch of the disk: this is usually referred as to a one-zone model.
To take account of the low ionization fraction of the disk, the induction
 equation includes both ohmic dissipation and the Hall effect.
We induce a transition from outburst to quiescence by an instantaneous
 decrease of the temperature.
The evolution of the MRI during the transition is found to be very sensitive
 to the temperature of the quiescent disk.
As long as the temperature is higher than a critical value of about 2000 K,
 MHD turbulence and angular momentum transport is sustained by the MRI.
However, MHD turbulence dies away within an orbital
 time if the temperature falls below this critical value.
In this case, the stress drops off by more than 2 orders of magnitude,
 and is dominated by the Reynolds stress associated with the remnant motions
 from the outburst.
The critical temperature depends
 slightly on the distance from the central star and the local density of
 the disk.
\end{abstract}


\section{INTRODUCTION}

Dwarf novae are close binary systems that consist of a white dwarf and a
Roche-lobe-filling secondary star.
The matter overflowing from the secondary star forms an accretion disk
around the white dwarf.
Dwarf nova systems show repetitive outbursts in which the luminosity of
the disks increase rapidly.
The duration of the outbursts ranges from a few to 20 days, and the
recurrence time is about 20 -- 300 days.
The disk instability model (Osaki 1974),
in which the evolution of the disk is regulated by changes in the rate of
angular momentum transport,
 is generally accepted as the
explanation of the outbursts (Cannizzo 1993; and references therein).
Since ordinary molecular viscosity is too low to explain the
evolutionary timescale of the disks, an anomalous stress associated with
turbulence and characterized by the $\alpha$ parameter (Shakura \&
Sunyaev 1973), is thought to be the source of angular momentum transport.
Detailed comparisons between observed light curves of dwarf novae and
the theoretical disk models suggest that the viscous parameter $\alpha$
must vary between the hot (outburst) and cold (quiescent) state, and the
amplitude of $\alpha_{\rm hot}$ and $\alpha_{\rm cold}$ is typically of
the order of 0.1 and 0.01, respectively.
However, while the origin of the anomalous stress has become better understood
 in recent years (see below), the reasons for the
difference between $\alpha_{\rm hot}$ and $\alpha_{\rm cold}$ are still
unclear.

The magnetorotational instability (MRI; Balbus \& Hawley 1991) is the
most promising source of the anomalous stress.
MHD turbulence driven by the MRI can transport angular momentum by
the Maxwell (magnetic) stress.
During the hot state, the disk gas is fully ionized so that the ideal
MHD approximation is appropriate.
Numerical simulations that adopt ideal MHD have shown that the Maxwell
stress caused by the MRI can give $\alpha \sim 0.1$ to 0.01, where the
saturation level is determined by the geometry and strength
of the magnetic field (e.g., Hawley, Gammie, \& Balbus 1995; 1996).
Therefore, MHD turbulence can be the primary mechanism of angular momentum
transport at least during outbursts, that is it can account for
$\alpha_{\rm hot}$.

During the cold state, on the other hand, the gas is only weakly ionized.
Gammie \& Menou (1998) have pointed out that ohmic dissipation could
modify the nature of the MHD turbulence at quiescence, and this could
make the difference between $\alpha_{\rm hot}$ and $\alpha_{\rm cold}$.
Calculations of the ionization fraction in dwarf nova disks
reveal that the Hall effect as well as ohmic dissipation must be
considered if the temperature is $T \lesssim 2000$ K (Sano \& Stone
2002a).
At low temperatures, ohmic dissipation can suppress the growth
of the MRI (Jin 1996; Sano \& Miyama 1999).
If turbulence dies away completely as a result of ohmic dissipation, the
Maxwell stress cannot provide the required anomalous stress during
quiescence (Menou 2000).
Thus, it is important to calculate numerically the amplitude of the
Maxwell stress at the cold state.

In this paper, we examine the behavior of MHD turbulence during the
transition from an outburst to quiescence using local three-dimensional
MHD simulations.  These simulations include the dominant non-ideal MHD
effects as determined from a self-consistent calculation of the
ionization state in the disk.  Since we adopt a local approximation to
study the turbulent stresses (usually referred to as a one-zone model,
e.g., Mineshige \& Osaki 1983; Cannizzo \& Wheeler 1984), our
calculations do not include global effects (such as spiral shocks in
the disk, Sawada, Matsuda, \& Hachisu 1986) which may contribute to
angular momentum transport.

The plan of this paper is as follows.  In \S~2, we calculate the
ionization fraction in dwarf nova disks by solving the Saha equation to
estimate the magnitude of nonideal MHD effects at quiescence.  Our
numerical method is described in \S~3.  The results of numerical
simulations which examine the dependence of the stress on the geometry
of magnetic field and the decay timescale of turbulence are presented
in \S~4.  In \S~5, we discuss the activity of MHD turbulence and the
source of angular momentum transport during quiescence.

\section{IONIZATION STATE OF THE DISKS}

During outburst, the temperature of the disk is above $10^4$
K, so that the gas is fully ionized and the ideal MHD approximation holds.
In this case, the disk will be dominated by MHD turbulence driven by the MRI,
with the plasma beta $\beta \sim
10^2$ -- $10^4$ (e.g., Hawley, Gammie, \& Balbus 1995; 1996), so that
the Alfv{\'e}n speed is typically $v_{\rm A} \sim 10^4$ --
$10^5$ cm~s$^{-1}$.
During the transition to quiescence, the temperature drops to
a few thousands K.
The ionization fraction decreases dramatically, and thus nonideal
MHD effects must be included.
The induction equation is then
\begin{equation}
\frac{\partial \mbox{\boldmath $B$}}{\partial t} =
\mbox{\boldmath $\nabla$} \times
\left[
\mbox{\boldmath $v$} \times \mbox{\boldmath $B$}
- \frac{4 \pi  \eta \mbox{\boldmath $J$}}{c}
- \frac{\mbox{\boldmath $J$} \times \mbox{\boldmath $B$}}{e n_e}
+ \frac{\left(\mbox{\boldmath $J$} \times \mbox{\boldmath $B$}\right)
\times \mbox{\boldmath $B$}}{c \gamma_i \rho_i \rho}
\right] ~,
\label{eqn:faraday}
\end{equation}
where {\boldmath $v$} is the neutral velocity, $\eta$ is the magnetic
diffusivity, {\boldmath $J$} is the current density, $c$ is the speed of
light, $e$ is the electron charge, $n_e$ is the number density of
electrons, $\gamma_i$ is the drag coefficient, and $\rho_i$ and $\rho$ are
the mass densities of ions and neutrals, respectively.
Each term in the brackets of equation (\ref{eqn:faraday}) from left to
right means the inductive term ($I$), ohmic dissipation ($O$), Hall
effect ($H$), and ambipolar diffusion ($A$). 

The importance of the nonideal effects can be measured by the relative
ratio of each term in equation (1) to the inductive term.
These ratios are determined mainly by the ionization
fraction.
At a typical density of dwarf nova disks ($n_n \sim 10^{18}$
cm$^{-3}$, where $n_n$ is the number density of neutrals),
ohmic dissipation and Hall effect are very important and affect the
evolution of the MRI (Sano \& Stone 2002a).
The size of the ohmic dissipation is given by the magnetic Reynolds
number;
\begin{equation}
Re_{M} = \frac{v_{\rm A}^2}{\eta \Omega} ~,
\label{eqn:rm}
\end{equation}
where $\Omega = ( G M / r^3)^{-1/2}$ is the angular velocity, $G$ is the
gravitational constant, $M$ is the mass of the central object, and $r$
is the distance from the central star.
The magnetic diffusivity is given by
\begin{equation}
\eta = \frac{c^2 m_e \nu_e}{4 \pi e^2 n_e} =
234 \left( \frac{n_e}{n_n} \right)^{-1} T^{1/2} ~{\rm cm}^2 {\rm s}^{-1},
\end{equation}
where $\nu_e = n_n \langle \sigma v \rangle_e$ is an effective collision
frequency for electrons.
Here we take $\langle \sigma v \rangle_e = 8.28 \times 10^{-10}
T^{1/2}$ (Draine, Roberge, \& Dalgarno 1983).
The magnetic Reynolds number $Re_{M}$ is the ratio of the inductive term
to the ohmic dissipation ($Re_{M} = I/O$; Sano \& Stone 2002a), and a
smaller $Re_{M}$ represents a more resistive case.
The Hall effect on the MRI is characterized by the Hall parameter
(Balbus \& Terquem 2001);
\begin{equation}
X = \frac{c B \Omega}{2 \pi e n_e v_{\rm A}^2} ~,
\end{equation}
which is proportional to the ratio between the Hall and the inductive
term ($X = 2 H/I$; Sano \& Stone 2002a).
If the Hall parameter $X$ is large, the Hall term is important.

Both the magnetic Reynolds number and Hall parameter can be written as a
function of the number density of electrons and the temperature;
\begin{equation}
Re_{M} = 68
\left( \frac{v_{\rm A}}{10^5 ~{\rm cm}~{\rm s}^{-1}} \right)^2
\left( \frac{n_e/n_n}{10^{-6}} \right)
\left( \frac{T}{3000 ~{\rm K}} \right)^{-1/2}
\left( \frac{M}{M_{\odot}} \right)^{-1/2}
\left( \frac{r}{10^{10} ~{\rm cm}} \right)^{3/2} ~,
\label{eqn:rem}
\end{equation}
and
\begin{equation}
X = 0.0059
\left( \frac{v_{\rm A}}{10^5 ~{\rm cm}~{\rm s}^{-1}} \right)^{-1}
\left( \frac{n_e/n_n}{10^{-6}} \right)^{-1}
\left( \frac{n_n}{10^{18} ~{\rm cm}^{-3}} \right)^{-1/2}
\left( \frac{M}{M_{\odot}} \right)^{1/2}
\left( \frac{r}{10^{10} ~{\rm cm}} \right)^{-3/2} ~.
\label{eqn:x}
\end{equation}
Interestingly, the product $Re_{M}X$ is independent of the ionization
fraction;
\begin{equation}
Re_{M}X = 0.40
\left( \frac{v_{\rm A}}{10^5 ~{\rm cm}~{\rm s}^{-1}} \right)
\left( \frac{n_n}{10^{18} ~{\rm cm}^{-3}} \right)^{-1/2}
\left( \frac{T}{3000 ~{\rm K}} \right)^{-1/2} ~.
\end{equation}

The number density of electrons is a function of the temperature,
because the free electrons in dwarf nova disks are mainly generated
by thermal ionization of Al and K.
Figure~\ref{fig:tdep-n18} shows the temperature dependence of the
electron abundance $n_e/n_n$.
We have solved the Saha equation assuming Solar abundance
for the case of $n_n = 10^{18}$ cm$^{-3}$, which is a typical
density of dwarf nova disks.
The ionization fraction decreases as the temperature decreases.
As seen from the figure, the fraction is very sensitive to the
temperature when $T \lesssim$ 2000~K.
For example, the abundance is $n_e/n_n \approx 3.5 \times 10^{-8}$ at $T =
2000$ K, but it is reduced by 2 orders of magnitude ($n_e/n_n \approx
3.7 \times 10^{-10}$) at $T = 1500$ K.

Figure~\ref{fig:tdep-n18} also shows the temperature dependence of the
parameters $Re_{M}$ and $X$, which are given by equations
(\ref{eqn:rem}) and (\ref{eqn:x}), respectively.
Here we assume $v_{\rm A} = 10^5$ cm~s$^{-1}$, $M = M_{\odot}$, and $r =
10^{10}$ cm.
When $Re_{M} \lesssim 1$ or $X \gtrsim 1$, these nonideal MHD terms can
modify the linear growth and nonlinear evolution of the MRI%
\footnote{The definition of the magnetic Reynolds number given by
equation~(\protect{\ref{eqn:rm}}) is different from that used in the
previous work
($Re_{M}' = c_s H / \eta$; Gammie \& Menou 1998; Menou 2000), and thus
the critical value is also different (see Sano \& Stone 2002a,b).}
(Sano, Inutsuka, \& Miyama 1998; Wardle 1999; Fleming, Stone, \& Hawley
2000; Balbus \& Terquem 2000; Sano \& Stone 2002a,b).
Then, nonideal MHD effects are clearly inefficient if the
temperature is higher than 3000~K ($Re_{M} \gg 1$ and $X \ll 1$), so
that ideal MHD may be valid.
However, both the ohmic dissipation and Hall effect become important
when $T \lesssim$ 2000~K ($Re_{M} \lesssim 1$ and $X \gtrsim 1$).
Therefore, these nonideal MHD effects must be included in the analysis
when the temperature of the disk becomes below 2000 K.
At this range of the temperature, the electron abundance and number
density is $n_e/n_n \lesssim 10^{-8}$ and $n_e \lesssim 10^{12} ~{\rm
cm}^{-3}$, respectively.

For a typical condition of dwarf nova disks, the relative ratio of the
ambipolar diffusion to the inductive term is given by
\begin{equation}
\frac{A}{I} = \frac{\Omega}{\gamma_i \rho_i} = 6.1 \times 10^{-6}
\left( \frac{n_i}{10^{12} ~{\rm cm}^{-3}} \right)^{-1}
\left( \frac{M}{M_{\odot}} \right)^{1/2}
\left( \frac{r}{10^{10} ~{\rm cm}} \right)^{-3/2} ~,
\label{eqn:ad}
\end{equation}
where we assume that $\gamma_i = \langle \sigma v \rangle_i / (m_i + m_n)$,
$\langle \sigma v \rangle_i = 1.9 \times 10^{-9}$ cm$^3$ (Draine et
al. 1983), the ion and neutral particle mass are $m_i =
30 m_{\rm H}$ and $m_n = 1.27 m_{\rm H}$, and $m_{\rm H}$ is the mass of
a hydrogen atom.
As mentioned in Gammie and Menou (1998), this ratio given by equation
(\ref{eqn:ad}) is so small that the ambipolar diffusion does not affect
the evolution of the MRI (Blaes \& Balbus 1994; Brandenburg et al. 1995;
Hawley \& Stone 1998).
Therefore, we ignore the ambipolar diffusion in the numerical
simulations in this paper.

\section{NUMERICAL METHOD}

We study the evolution of the MRI and MHD turbulence during the transition
to quiescence using three-dimensional MHD simulations.
The shearing box approximation (Hawley et al. 1995) is
adopted to investigate the local evolution of the instability.
The equations of adiabatic MHD with the ratio of specific heats
$\gamma = 5/3$ are solved in a Cartesian frame of reference ($x$, $y$,
$z$) corotating with the disk at the angular velocity $\Omega$, where
$x$ is the radial, $y$ is the azimuthal, and $z$ is the vertical
direction.
The induction equation includes terms that describe ohmic dissipation
and the Hall effect.
The details of the equations and our numerical scheme are described in
Sano \& Stone (2002b).
Vertical gravity is ignored, so that the calculations
begin with a uniform density, $\rho = \rho_0$, and a uniform pressure,
$P = P_0$.
The magnetic field structure of dwarf nova disks is still highly
uncertain.  If the white dwarf has a strong magnetic field, then the
dipole field may penetrate the disk vertically.  On the other hand, a 
toroidal field amplified by the differential rotation may be dominant in 
the disk.  In this case, the net flux of the vertical field could be
negligible.  Nonlinear evolution of the MRI started with a toroidal
field is quite similar to that with a zero net flux $B_z$ (Sano \& Stone 
2002b).  Therefore, in this paper, we consider two kinds of initial
field geometries; a uniform vertical field, $B_z = B_0$, and a zero net
flux vertical field, $B_z = B_0 \sin (2 \pi x / H)$, where $H =
(2/\gamma)^{1/2} c_{s0} / \Omega$ is the scale height of the disk and
$c_{s0}$ is the initial sound speed.
We choose normalizations with $\rho_0 = 1$, $H = 1$, and $\Omega =
10^{-3}$.
The magnetic field strength is characterized by the plasma beta
$\beta_0 = P_0 / (B_0^2/8 \pi) = 3200$.
This is about 100 G in dwarf nova disks, which corresponds to a dipole
field of a white dwarf with a moderately strong magnetic
field ($\sim 10^6$ G) at a radius $r = 10^{10}$ cm.
The size of the simulation box is $H \times 4H \times H$ for all the
models calculated in this paper.
The initial ratio of the MRI wavelength, $2 \pi v_{{\rm
A}0}/\Omega$, to the box height is about 0.11.
A standard grid resolution of $32 \times 128 \times 32$ uniform zones is
used for most of our calculations.

To set up a disk in outburst, the induction equation is integrated for 50
orbits assuming ideal MHD.
The simulations are then restarted with nonideal MHD effects included in
the induction equation, and the models are evolved for a further 50
orbits.
Although the transition should proceed at the thermal timescale of the
disk, our technique assumes an instantaneous change of the temperature
at 50 orbits, which is thereafter kept constant.
The quiescent disk is characterized by the
non-dimensional parameters $Re_{M}$ and $X$.
We have computed 4 different models with $(Re_M, X) = (100,0.01)$,
(10,0.1), (1,1), and (0.1,10), which correspond to the temperature $T =
2800$, 2200, 1800, and 1600 K, respectively (see Fig. 1).

As seen from equations (\ref{eqn:rem}) and (\ref{eqn:x}), the parameters
$Re_M$ and $X$ are functions of the magnetic field strength.
Because the vertical component of the field is essential for the
evolution of the MRI (Sano \& Stone 2002b), we define these parameters
($Re_{M,50}$ and $X_{50}$) using the volume-averaged
vertical field strength $\langle B_z^2 \rangle^{1/2}$ and squared Alfv{\'e}n
speed $\langle v_{{\rm A}z}^2 \rangle$ at 50 orbits, i.e., $Re_{M,50} =
\langle v_{{\rm A}z}^2 \rangle / \eta \Omega$ and $X_{50} = c \langle
B_z^2 \rangle^{1/2} \Omega / 2 \pi e n_e \langle v_{{\rm A}z}^2 \rangle$.
Then the size of $\eta$ and $n_e$ after 50 orbits are obtained from
$Re_{M,50}$ and $X_{50}$, respectively.
For simplicity, the electron abundance $n_e/n_n$ is assumed to be
constant in time and uniform in space throughout the calculation.
Since the pressure has little effect on the evolution of the MRI, we
find it does not matter whether or not the pressure is changed to be
consistent with the reduced temperature of the disk after 50 orbits.
We have not changed the pressure in the models reported here.

\section{NUMERICAL RESULTS}

\subsection{Models without a Net Flux of the Magnetic Field}

First, models started with a zero net flux $B_z$ are
studied in this subsection.
All the calculated models are listed in Table~\ref{tbl:s}.
To obtain MHD turbulence in the hot state, model S
is calculated for 50 orbits using the ideal MHD equations.
The models of quiescent disk are labeled SA, SB, SC, and SD.
For each model, ohmic dissipation and the Hall term are included after 50 orbits.
The size of the parameters $Re_{M,50}$ and $X_{50}$ are listed in the
Table.
The magnetic diffusivity and the electron abundance at quiescence
are calculated using the magnetic field strength of model S at 50 orbits.
The turbulent stress is the sum of the Maxwell stress, $w_M = - B_x B_y
/ 4 \pi$, and the Reynolds stress, $w_R = \rho v_x \delta v_y$, where
$\delta v_y$ is the perturbed azimuthal velocity from the Keplerian
flow.
The $\alpha$ parameter of Shakura \& Sunyaev (1973) is given by
$\alpha = (w_M + w_R)/P_0$.
The time- and volume-averaged stresses ($\langle \negthinspace \langle
w_M \rangle \negthinspace \rangle$ and $\langle \negthinspace \langle
w_R \rangle \negthinspace \rangle$) are listed in Table~\ref{tbl:s}.
The time averages are taken from 75 to 100 orbits.

Figure~\ref{fig:wm-t50-s} shows the time evolution of the
volume-averaged Maxwell stress $\langle w_M \rangle / P_0$,
where time is measured in orbits $t_{\rm rot} = 2 \pi / \Omega$.
The ideal MHD model (S) is shown by the dashed curve.
The saturation level of the Maxwell stress in model S is of the
order of 0.01 throughout the evolution.
The solid curves after 50 orbits are for the models that include the
nonideal MHD effects.
From top to bottom, they are models SA ($T = 2800$ ~K), SB (2200 K), SC
(1800 K), and SD (1600 K).

The suppression of the MRI is mainly caused by ohmic dissipation
(Sano \& Stone 2002b).
When ohmic dissipation is inefficient ($Re_{M,50} > 100$), the
amplitude of the MHD turbulence maintains a level similar to the
ideal MHD case.
At $Re_{M,50} = 10$, the Maxwell stress is reduced by a factor of 4, but
still has a significant amplitude ($\langle \negthinspace \langle w_{M}
\rangle \negthinspace \rangle / P_0 = 4.7 \times 10^{-3}$).
However, if $Re_{M,50} \lesssim 1$ or $T \lesssim 1800$ K, the MHD
turbulence dies away very quickly.
As the temperature decreases, the amplitude of the magnetic stress
decreases dramatically.
For the model SD, the magnetic field and stress are almost vanished at
the end of the calculation, i.e., 50 orbits after the transition.
The duration of quiescence in dwarf novae is typically more than a few
tens of days, which corresponds to over 1000 orbits.
Thus, the magnetic field in the disks can disappear locally during
quiescence if the temperature is less than about 1500 K.

Figure~\ref{fig:w-rm-s} shows the saturation level of the Maxwell
$\langle \negthinspace \langle w_M \rangle \negthinspace \rangle$ and
Reynolds stress $\langle \negthinspace \langle w_R \rangle \negthinspace
\rangle$ as a function of the magnetic Reynolds number $Re_{M,50}$.
The stresses are normalized in terms of the initial pressure $P_0$, so
that they have the same dimension as the $\alpha$ parameter.
The upper and lower arrows indicate the Maxwell and Reynolds stress for
the ideal MHD model ($Re_{M} \rightarrow \infty$).
As shown in Figure~\ref{fig:wm-t50-s}, active MHD turbulence can be
sustain by the MRI if the magnetic Reynolds number is much larger than
unity.
Angular momentum transport is dominated by the Maxwell stress, which
is about 4 times larger than the Reynolds stress.

When $Re_{M,50} \lesssim 1$, the MRI is suppressed by the ohmic
dissipation and MHD turbulence dies away immediately.
The saturation level of the Maxwell stress is quite small ($\langle
\negthinspace \langle w_M \rangle \negthinspace \rangle / P_0 \lesssim
10^{-6}$) compared with the less resistive runs (S, SA, and SB).
This dependence on the magnetic Reynolds number is consistent with
the results of previous work (Fleming et al 2000; Sano \& Stone 2002b).
While the Maxwell stress decreases as the temperature decreases,
it is found that the Reynolds stress has an amplitude which is
comparable to or larger than the magnetic stress.
Since we do not include an explicit kinematic viscosity in our calculations,
and since the numerical viscosity of our algorithm is
much lower than the explicit magnetic
diffusivity added to model quiescence, epicyclic oscillations
remain even after the magnetic field disappears.
Although the amplitude of the Reynolds stress is very small ($\alpha
\approx \langle \negthinspace \langle w_{R} \rangle \negthinspace
\rangle / P_0 \sim 10^{-6}$), these motions, which are remnants of the
turbulence during outburst,
could give non zero accretion stress during quiescence.
However, such motions may be swamped by global effects which cannot be
studied in our local simulations.

The reduction of the Maxwell stress due to the nonideal MHD effects is
independent of the numerical resolution.
We have performed the same simulations as shown by Figures~\ref{fig:tdep-n18}
-- \ref{fig:w-rm-s} with double the resolution ($64 \times 256 \times 64$).
The saturation level of the stress in the high-resolution runs are
listed in Table~\ref{tbl:s}.
For the ideal MHD case, the Maxwell stress in the high-resolution run is
in fact slightly smaller than that in the standard run.
For less resistive models (SA and SB), the Maxwell stress is always
larger than the Reynolds stress and the saturation level of the stress
is almost unrelated with the magnetic Reynolds number.
When $Re_{M} \lesssim 1$, on the other hand, the ohmic dissipation
suppresses the MRI and MHD turbulence.
The Maxwell stress is $\langle \negthinspace \langle w_M \rangle
\negthinspace \rangle / P_0 = 1.0 \times 10^{-6}$ and $1.0 \times
10^{-10}$ for model SC and SD, respectively.
For these cases, angular momentum transport is dominated by the
Reynolds stress with a similar amplitude to the standard-resolution
runs ($\alpha \sim 10^{-6}$).

\subsection{Models with a Net Flux of the Vertical Field}

Next we consider the effect of the initial field geometry.
Models started with a uniform $B_z$ are listed in Table~\ref{tbl:z},
along with the model parameters and saturation level of the
Maxwell and Reynolds stress.
With this field geometry, the saturation level of $\alpha$ in the ideal
MHD run (Z) is of the order of 0.1.
This is an order of magnitude larger than that in the zero net flux $B_z$
model (S).
For models with a net flux of vertical field, a two-channel
flow forms repeatedly at the nonlinear phase of the MRI (Sano \&
Inutsuka 2001).
The channel flow is destroyed through magnetic reconnection
triggered by the parasitic instability (Goodman \& Xu 1994).
Since the magnetic energy and stress increase exponentially when the
channel flow is growing, the time evolution of the magnetic stress shows
frequent spike-shaped excursions.
The growth of the two-channel flow can be seen only in the uniform $B_z$
runs, and this causes the relatively larger magnetic stress compared
with the zero net flux $B_z$ runs (Sano \& Stone 2002b).

For models of a quiescent disk (ZA, ZB, ZC, and ZD), we choose
the same sets of $Re_{M,50}$ and $X_{50}$ as for the zero net flux $B_z$
runs.
In Figure~\ref{fig:w-rm-z}, the Maxwell and Reynolds stress are depicted
by the filled and open circles, respectively.
The dependence of the stresses on the size of the nonideal effects is
qualitatively quite similar to that for the zero net flux $B_z$ runs.
When the magnetic Reynolds number is larger than 10, the stress is
almost independent of $Re_{M,50}$ and $X_{50}$.
The difference in the Maxwell stress between the ideal MHD (Z) and the
$Re_{M,50} = 10$ (ZB) runs is only 10 \%.
The Maxwell stress always dominates the Reynolds stress by a factor of 5.

If $Re_{M,50} \lesssim 1$, on the other hand, the Maxwell stress becomes
inefficient ($\langle \negthinspace \langle w_{M} \rangle \negthinspace
\rangle / P_0 \sim 10^{-6}$) as a result of the suppression of the MRI.
For models ZC and ZD, the spatial fluctuations in the magnetic field
disappear soon after 50 orbits and the field geometry returns to a
uniform $B_z$.
The $\alpha$ parameter is dominated by the Reynolds stress, and
the saturation level of $\langle \negthinspace \langle w_{R} \rangle
\negthinspace \rangle / P_0$ is larger than those in the zero net flux
$B_z$ models (SC and SD).
We also calculated a model with $Re_{M,50} = 0.01$ and $X_{50} = 100$
($T = 1400$ K).
In this case, the critical wavelength for the MRI is longer than the
scale height of the disk (i.e., the simulation box size), because
ohmic dissipation stabilizes the shorter-wavelength fluctuations.
The saturation level of the Maxwell and Reynolds stress are $\langle
\negthinspace \langle w_M \rangle \negthinspace \rangle / P_0 = - 3.6
\times 10^{-6}$ and $\langle \negthinspace \langle w_R \rangle
\negthinspace \rangle / P_0 = 2.6 \times 10^{-3}$, respectively.
The Reynolds stress still has the largest amplitude, but this
is a remnant of the turbulence during outburst.
Once again, we emphasize the amplitude of the Reynolds stress may be different
in global calculations.

\subsection{Decay Timescale of MHD Turbulence}

When the magnetic Reynolds number is less than unity, MHD turbulence
immediately dies away after the transition.
We define the decay time $t_{\rm dec}$ of the turbulence as the time it
takes for the magnetic energy to fall to 1/e of its value at 50 orbits.
The decay time for models with $Re_{M,50} = 1$ (SC and ZC) is
$t_{\rm dec} / t_{\rm rot} = 0.36$ and 0.12, where $t_{\rm rot}$ is the
orbital time.
The $Re_{M,50} = 0.1$ runs show much faster decay; $t_{\rm dec} /
t_{\rm rot} = 0.028$ and 0.0077 for models SD and ZD, respectively.
The decay time in the high-resolution runs is slightly shorter than that
in the standard runs ($t_{\rm dec} / t_{\rm rot} = 0.16$ for model SC
and $t_{\rm dec} / t_{\rm rot} = 0.012$ for model SD).
Thus all the decay times we obtained are less than an orbit.

The dissipation timescale of the magnetic field is given by $t_{\rm dis}
= L^2 / \eta$, where $L$ is a typical length scale.
The turbulent motion is driven most effectively at the most unstable
wavelength for the MRI ($\lambda \sim v_{\rm A} / \Omega$), which is
much smaller than the disk scale height.
We find that the decay times $t_{\rm dec}$ obtained by the numerical
simulations are equivalent to the dissipation timescale with $L = v_{\rm
A} / \Omega$;
\begin{equation}
\frac{t_{\rm dis}}{t_{\rm rot}} =
\frac{\langle v_{\rm A}^2 \rangle / \Omega^2}{\eta} \frac{\Omega}{2 \pi} =
0.16 Re_{M,50} ~.
\end{equation}
This relation is valid only when the magnetic Reynolds number is less
than unity.

\section{DISCUSSION}

\subsection{The Critical Temperature and Transition Timescale}

When the magnetic Reynolds number falls to unity, MHD turbulence is
suppressed by ohmic dissipation independent of the amplitude
of the Hall term (Sano \& Stone
2002a,b).
The critical temperature corresponding to $Re_{M} = 1$ is about
2000 K for our model.
As seen from equation~(\ref{eqn:rem}), however, the critical temperature
depends on the neutral density $n_n$ and the distance from the central
star $r$.
Figure~\ref{fig:tcrit} shows the critical temperature $T_{\rm crit}$ as
a function of $r$ for the cases of the neutral density $n_n = 10^{17}$,
$10^{18}$, and $10^{19}$ cm$^{-3}$.
We assume the Alfv{\'e}n speed $v_{A} = 10^5$ cm s$^{-1}$ and the mass
of the central star $M = M_{\odot}$.
At a given temperature, the electron abundance decreases as the neutral
density increases, so that
the critical temperature is slightly higher when the density is
higher.
The MRI wavelength ($\sim v_{\rm A} / \Omega$) is shorter at a smaller
radius, because the angular velocity is larger.
Therefore, the MRI can be suppressed with a smaller amount of
diffusivity $\eta$ in the inner regions of the disk.
The critical temperature is about 2500 and 1500 K at $r = 10^9$ and
$10^{11}$ cm, respectively.
Since the magnetic Reynolds number, or the electron abundance, has a
steep dependence on the temperature, the range of the critical
temperature is quite narrow and the difference is at most a
factor of 2 over the entire disk.

In our numerical analysis, we assume a fixed temperature during
quiescence.
Thus the electron abundance at quiescence is assumed to be constant.
In a real disk, the critical temperature could be a little higher due to
a runaway decay of MHD turbulence (Menou 2000).
The decrease of the temperature makes both the electron abundance lower and
nonideal MHD effects more important.
The turbulent stress is therefore suppressed further, so that the temperature falls
 even more.
However, a very weak dependence of the stress on the
temperature can be seen in our models when $T \gtrsim 3000$ K.
This indicates that the runaway temperature should be less then 3000 K.
Although a self-consistent study of the evolution of the temperature and
electron abundance is needed to obtain a real $T_{\rm crit}$,
Figure~\ref{fig:tcrit} may give a fairly good estimate of the critical
temperature in dwarf nova disks.

We also assume the transition to quiescence
occurs instantaneously.
In actual systems, however, this may proceed at the thermal timescale
$t_{\rm th}$, where $t_{\rm th}/t_{\rm rot} \sim 1 / \alpha
\gtrsim 100$ (Frank, King, \& Raine 1992).
Suppose the temperature decreases gradually.
As long as the temperature is higher than the critical
value, we have shown that the accretion stress is nearly constant.
But, once the temperature drops below this critical value, the MRI is
suppressed,  the turbulence decays within an orbit, and
the $\alpha$
parameter drops by more than 2 orders of magnitude.
Therefore, even though the decrease of the temperature is gradual, the
transition timescale of the stress may be very much shorter.

\subsection{Onset of the Next Outburst}

Dwarf nova systems undergo recurrent outbursts.
The matter from the secondary star accumulates in the disk during
quiescence.
When the surface density exceeds a critical value at some region in the
disk, the thermal instability sets in and the temperature goes up.
Here we examine the behavior of the MRI at the transition to the hot
state.

Figure~\ref{fig:w-t100-z} shows the time evolution of the Maxwell and
Reynolds stress for model ZC until 100 orbits.
After 100 orbits, the induction equation for the ideal MHD is used in
order to imitate the onset of the next outburst.
As seen from the figure, the growth of the MRI starts immediately at the
transition to the outburst and the magnetic field is amplified
exponentially.
In a few orbits, the Maxwell stress reaches the same saturation level
as that before 50 orbits.
The key to this evolution is the existence of a net vertical flux.
During quiescence (from 50 to 100 orbits), MHD turbulence is
suppressed, but the magnetic flux within the shearing box is
conserved throughout the evolution.
Thus the field geometry at 100
orbits is a nearly uniform vertical field.
When the nonideal MHD effects are turned off, the linear
growth of the MRI starts and MHD turbulence is initiated.

This evolution may be different if the system has no net magnetic flux.
For example, if the magnetic field is {\em completely} dissipated during
quiescence (e.g., model SD), MHD-driven activity cannot occur during the
next outburst.
However, if enough magnetic field remains for there to be unstable modes even in a
small region of the disk, MHD turbulence develops in that region and
finally may spread over a large part of the disk (Hawley \& Balbus 1991).
Therefore, it is quite important to understand the global structure of
the magnetic field during quiescence, including the
magnetosphere of the white dwarf and the magnetic field supplied by
accreting material from the secondary star (Meyer \& Meyer-Hofmeister
1999).

\subsection{Beyond the Local Model}

Since we focus on the local behavior of MHD turbulence in this paper, neither global
structural effects nor global instabilities have been considered.
In real disks, global effects such as spiral shocks
(Sawada et al. 1986) may also transport angular momentum.

From our numerical results, the dependence of the $\alpha$ parameter
associated with local turbulence on temperature in a dwarf nova disk
can be summarized as follows.  The contribution of the MRI to the
$\alpha$ parameter is significant during outburst.  When the
temperature is higher than the critical value $T_{\rm crit} \sim 2000$~
K, the accretion stress is dominated by the Maxwell stress, with
$\alpha \sim 0.1$ and 0.01 for the cases with and without net vertical
flux, respectively.  At quiescence, on the other hand, $\alpha$ is very
sensitive to the temperature of the disk.  If $T \lesssim T_{\rm
crit}$, the stress is more than 2 orders of magnitude smaller than that
during outburst.

There are several reasons why global effects may influence these
results.  Firstly, the presence of a net vertical flux can affect the
amplitude of the turbulence stress during outburst, that is $\alpha_{\rm
hot}$.  In agree with the previous works using the local shearing box
(e.g., Hawley et al. 1995; 1996), our simulations have shown that an
sufficient amount of angular momentum transport in the hot state
requires the existence of a net vertical field.  On the other hand,
recent global simulations obtained a higher stress ($\alpha \sim 0.1$)
even without a net magnetic flux (Stone \& Pringle 2001; Hawley, Balbus, 
\& Stone 2001).  There is no observational constraint yet in the
structure and strength of the magnetic field in dwarf nova disks.
Thus, theoretical modeling of the global structure of the magnetic field
is quite important for a more detailed analysis of $\alpha$ at the hot
state.
Secondly, the ionization fraction of the disk gas, and therefore the
importance of non-ideal MHD effects, is extremely sensitive to
temperature.  Thus global models in which the radial and vertical
temperature profiles in the disk are calculated self-consistently with
the local heating rate due to turbulent stresses and cooling rate due to 
radiation are required.  
Furthermore, at the surface of the disk, the radiation from a hot 
white dwarf could be important source of heating and ionization
(Hameury, Lasota, \& Dubus 1999; Menou 2002).
Finally, even when Ohmic dissipation completely suppresses the MRI, small
amplitude motions remain in the disk and can give a non-zero Reynolds
stress.  In the local model, these motions are damped on a viscous
timescale, which may be very long.  However, they may be affected by
global density waves in the disk or tidally induced spiral shocks.
Thus, global MHD simulations are an important next step for
understanding the dynamics of dwarf nova disks.

\acknowledgements
We would like to thank John Cannizzo, Kristen Menou, and Neal Turner for
helpful discussions.
Computations were carried out on the VPP5000 at the National Astronomical
Observatory of Japan, and the VPP700 at the Subaru Telescope, NAOJ.
This work was supported by grants from the NASA OSS program and the NSF.


\clearpage

\begin{deluxetable}{ccccllll}
\small
\tablecaption{Zero Net Flux $B_z$ Simulations}
\tablehead{
\colhead{} & \colhead{} & \colhead{} & \colhead{} &
\multicolumn{2}{c}{Standard Resolution} &
\multicolumn{2}{c}{High Resolution} \\
\cline{5-6} \cline{7-8}
\colhead{Model} &
\colhead{$Re_{M,50}$} & \colhead{$X_{50}$} & \colhead{$T$ [K]} &
\colhead{$\langle\negthinspace\langle w_M
\rangle\negthinspace\rangle/P_0$} &
\colhead{$\langle\negthinspace\langle w_R
\rangle\negthinspace\rangle/P_0$} &
\colhead{$\langle\negthinspace\langle w_M
\rangle\negthinspace\rangle/P_0$} &
\colhead{$\langle\negthinspace\langle w_R
\rangle\negthinspace\rangle/P_0$}
}
\startdata
S  & $\infty$ & 0 & $> 10^4$ &
                         0.0222 & 0.00545 & 0.0101 & 0.00269 \nl
SA & 100 & 0.01 & 2800 & 0.0171 & 0.00417 & 0.0101 & 0.00266 \nl
SB & 10  & 0.1  & 2200 & 0.00467 & 0.00113 & 0.00591 & 0.00160 \nl
SC & 1   & 1    & 1800 & $1.00 \times 10^{-6}$ & $1.04 \times 10^{-6}$ &
$3.23 \times 10^{-7}$ & $7.18 \times 10^{-7}$ \nl
SD & 0.1 & 10   & 1600 & $1.87 \times 10^{-10}$ & $1.41 \times 10^{-6}$ &
$- 3.99 \times 10^{-9}$ & $3.39 \times 10^{-6}$ \nl
\enddata
\label{tbl:s}
\end{deluxetable}

\begin{deluxetable}{ccccll}
\small
\tablecaption{Uniform $B_z$ Simulations}
\tablehead{
\colhead{Model} &
\colhead{$Re_{M,50}$} & \colhead{$X_{50}$} & \colhead{$T$ [K]} &
\colhead{$\langle\negthinspace\langle w_M
\rangle\negthinspace\rangle/P_0$} &
\colhead{$\langle\negthinspace\langle w_R
\rangle\negthinspace\rangle/P_0$}
}
\startdata
Z  & $\infty$ & 0 & $> 10^4$ &
                         0.0809 & 0.0166 \nl
ZA & 100 & 0.01 & 2800 & 0.0802 & 0.0160 \nl
ZB & 10  & 0.1  & 2200 & 0.0717 & 0.0138 \nl
ZC & 1   & 1    & 1800 & $8.61 \times 10^{-8}$ & $8.80 \times 10^{-6}$ \nl
ZD & 0.1 & 10   & 1600 & $1.25 \times 10^{-6}$ & $3.98 \times 10^{-4}$ \nl
\enddata
\label{tbl:z}
\end{deluxetable}

\clearpage

\plotone{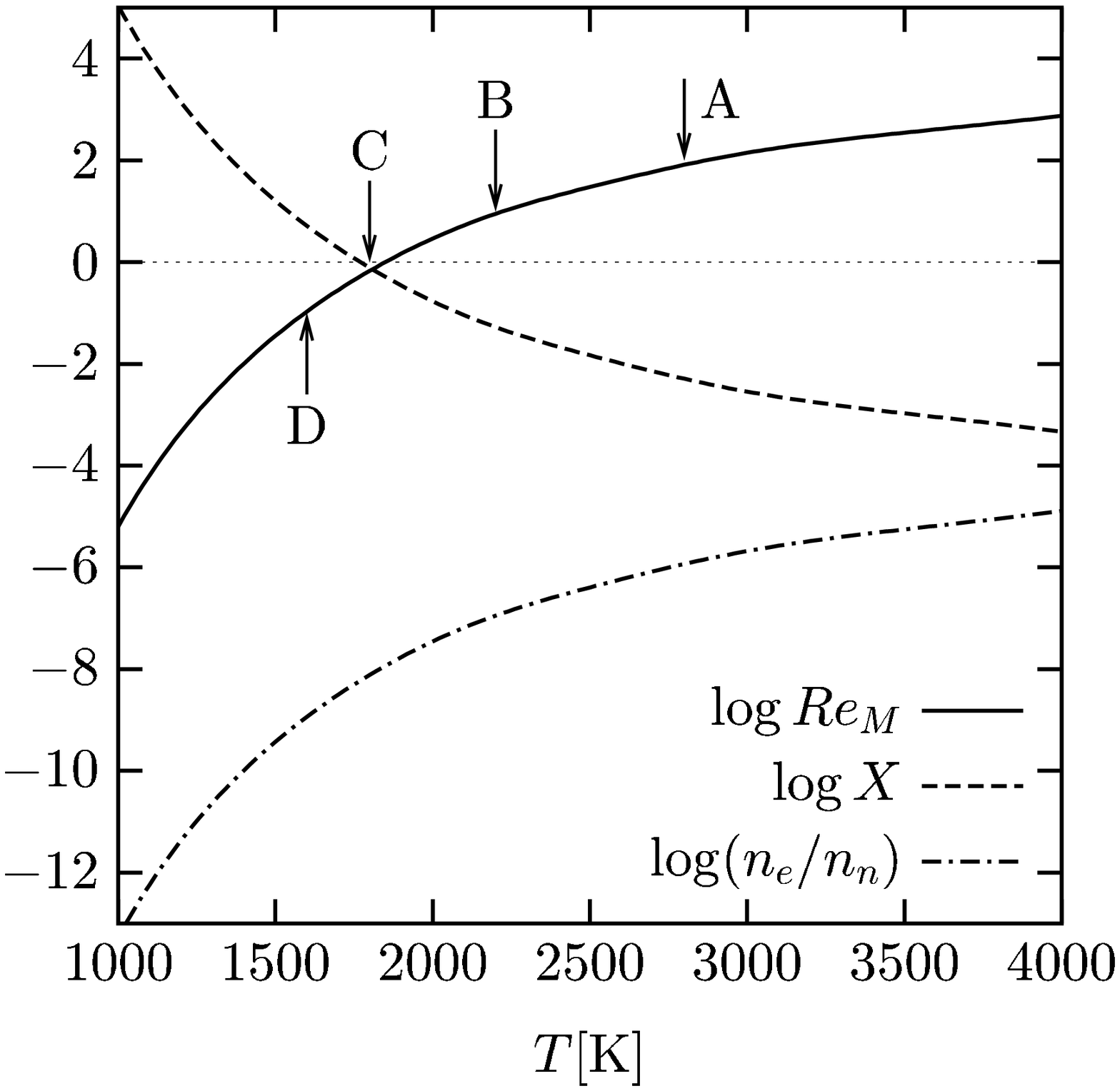}
\figcaption[fig1.ps]
{Temperature dependence of the electron abundance $n_e/n_n$ ({\it
dot-dashed curve}), the magnetic Reynolds number $Re_{M}$ ({\it solid
curve}), and the Hall parameter $X$ ({\it dashed curve}).
Typical quantities for dwarf nova disks are assumed in the calculations
of these curves; $n_n = 10^{18}$ cm$^{-3}$, $M = M_{\odot}$, $r = 10^{10}$
cm, and $v_{\rm A} = 10^5$ cm s$^{-1}$.
Arrows denotes the loci of our 4 models for quiescent disk (see
Tables~\protect{\ref{tbl:s}} and \protect{\ref{tbl:z}}).
\label{fig:tdep-n18}}

\newpage
\plotone{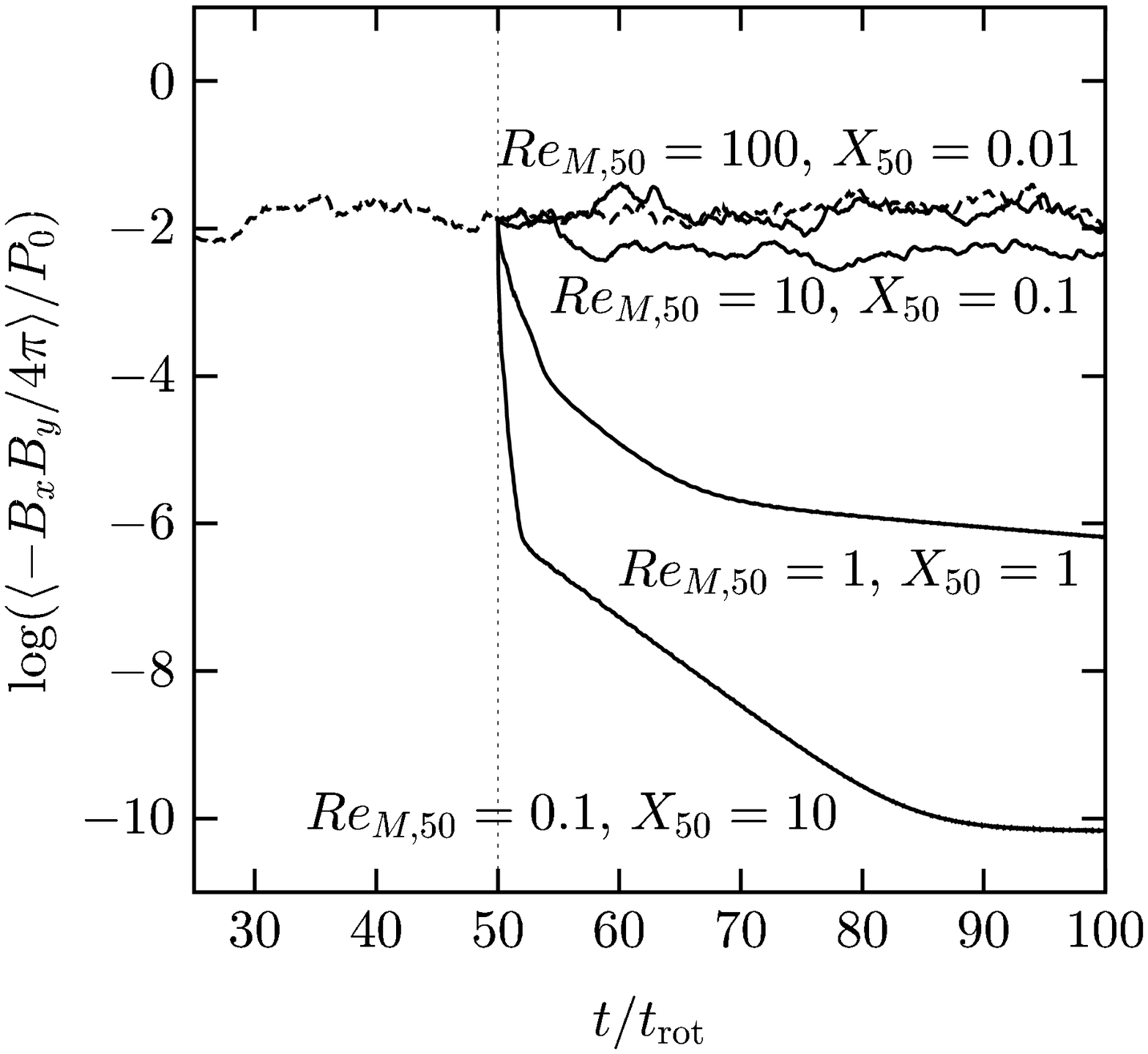}
\figcaption[fig2.ps]
{Time evolution of the Maxwell stress for the ideal MHD model ({\it
dashed curve}) and the models including ohmic dissipation and the Hall
effect ({\it solid curves}).
Nonideal MHD terms are included after 50 orbits, and the size of the terms
for each model is indicated in the figure.
\label{fig:wm-t50-s}}

\newpage
\plotone{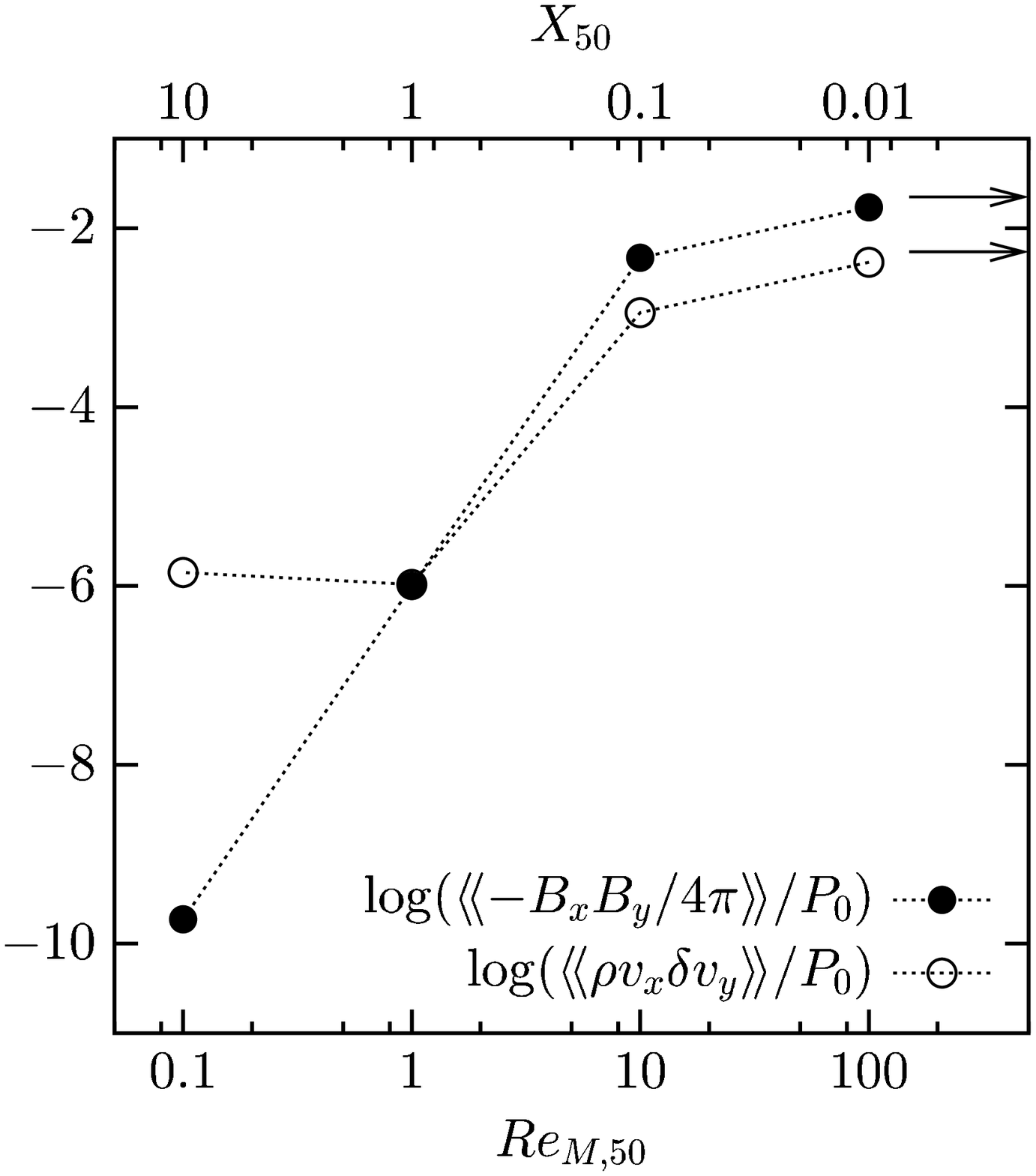}
\figcaption[fig3.ps]
{Time- and Volume-averaged Maxwell stress and Reynolds stress as a
function of the magnetic Reynolds number at 50 orbits ($Re_{M,50}$) for
zero net flux $B_z$ runs.
The Hall parameter at 50 orbits ($X_{50}$) is also shown at the top of
the figure.
The time average is taken from 75 to 100 orbits.
\label{fig:w-rm-s}}

\newpage
\plotone{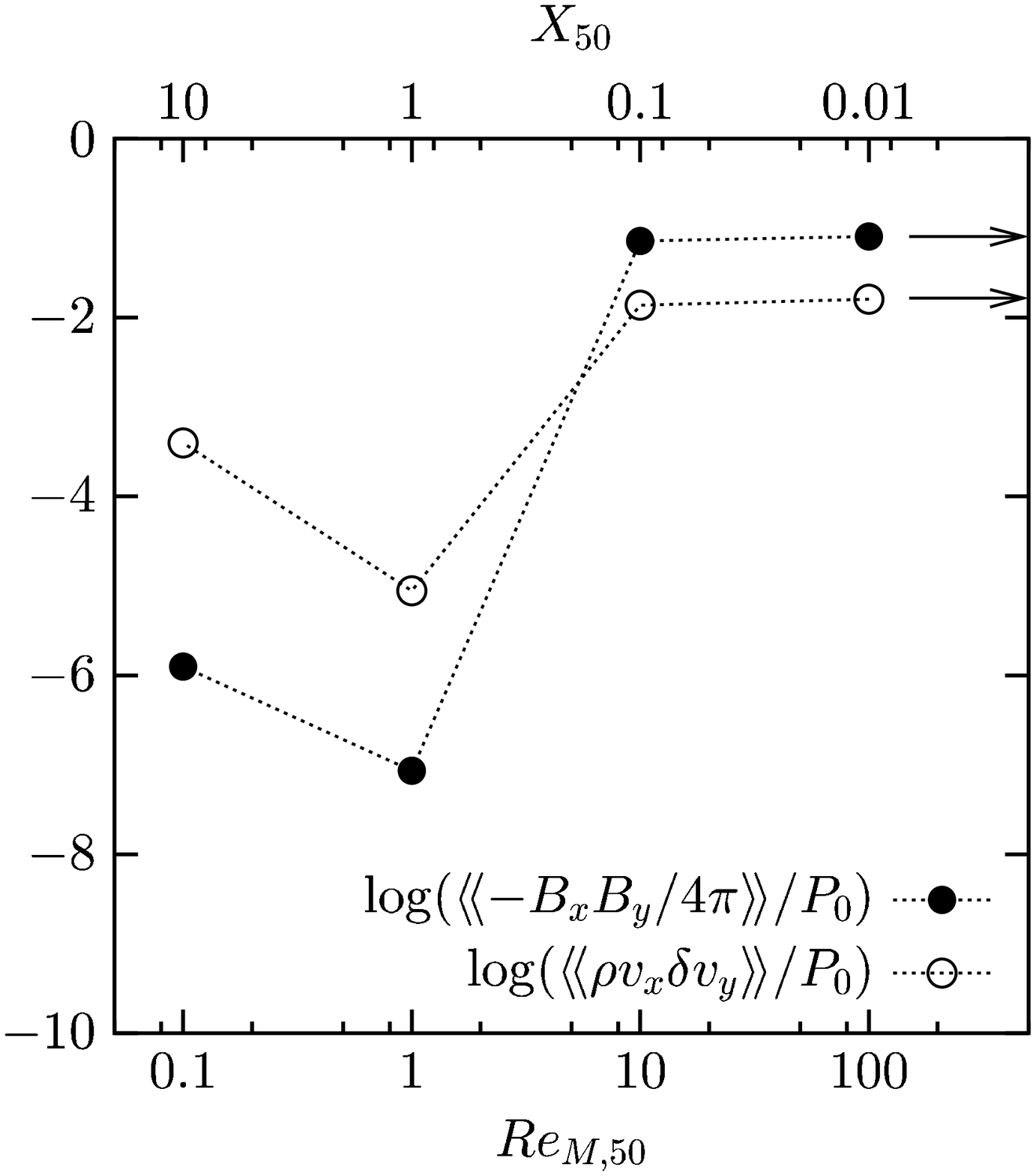}
\figcaption[fig4.ps]
{Time- and Volume-averaged Maxwell stress and Reynolds stress as a
function of the magnetic Reynolds number at 50 orbits ($Re_{M,50}$) for
uniform $B_z$ runs.
The Hall parameter at 50 orbits ($X_{50}$) is also shown at the top of
the figure.
The time average is taken from 75 to 100 orbits.
\label{fig:w-rm-z}}

\newpage
\plotone{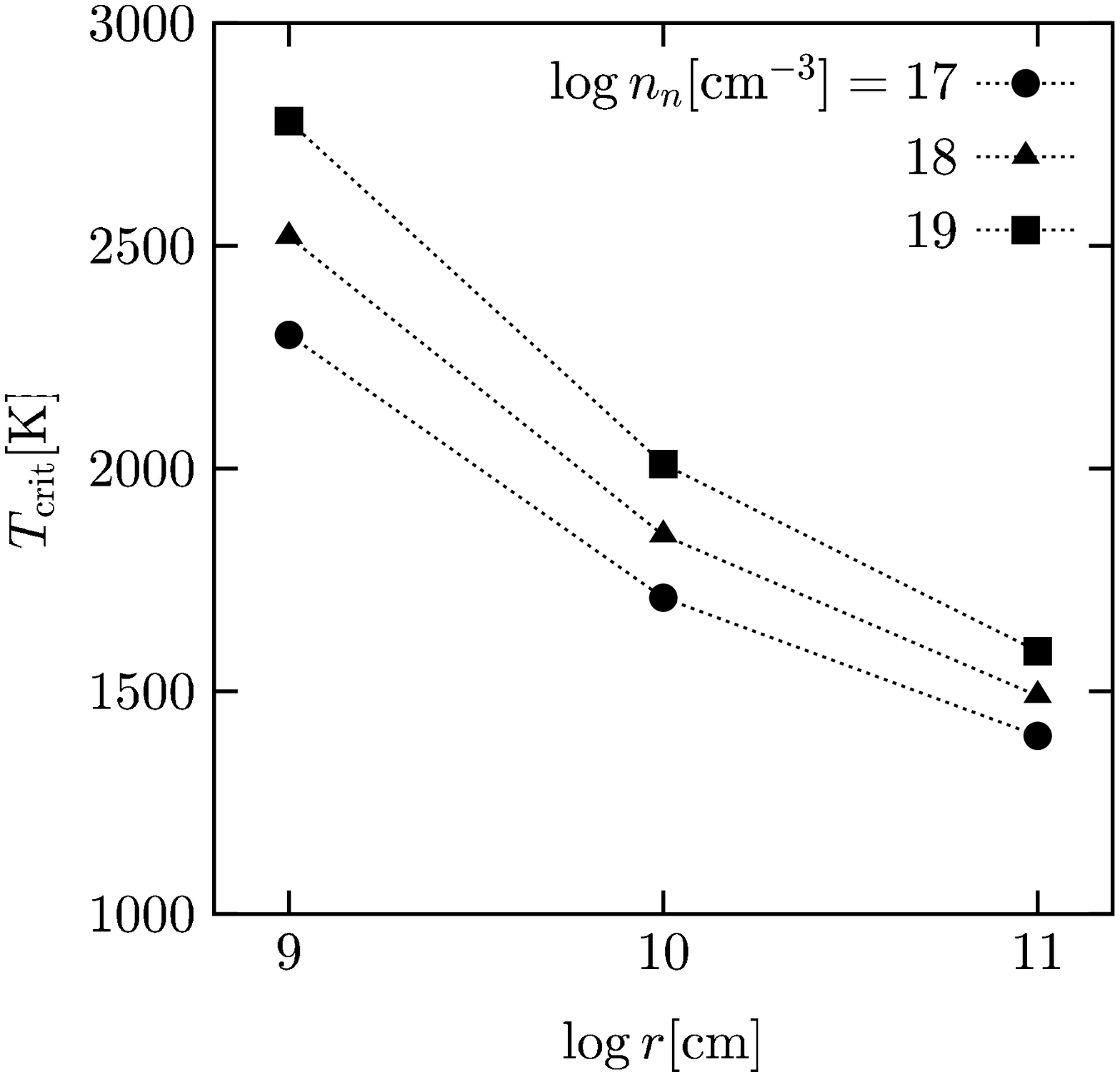}
\figcaption[fig5.ps]
{The critical temperature $T_{\rm crit}$ [K], at which the magnetic
Reynolds number is unity, is shown as a function of the distance from
the central star $r$ [cm] for the cases of the neutral density $n_n =
10^{17}$ cm$^{-3}$ ($circles$), $10^{18}$ ($triangles$), and
$10^{19}$ ($squares$).
The Alfv{\'e}n speed and the mass of the central star are assumed to be
$v_{\rm A} = 10^5$ cm s$^{-1}$ and $M = M_{\odot}$, respectively.
\label{fig:tcrit}}

\newpage
\plotone{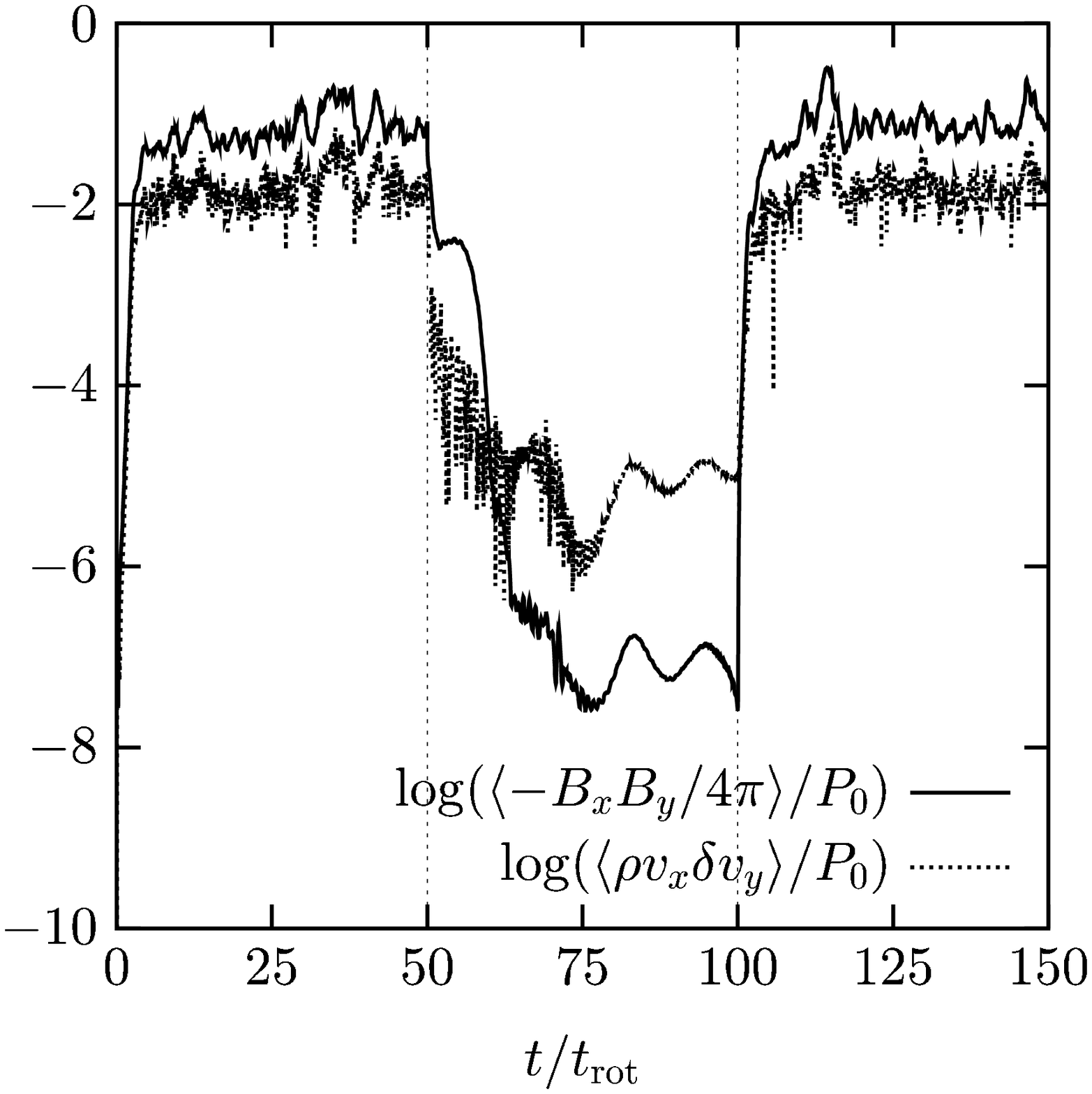}
\figcaption[fig6.ps]
{Time evolution of the Maxwell stress and Reynolds stress for model ZC
($Re_{M,50} = 1$ and $X_{50} = 1$).
After 100 orbits, the ideal MHD approximation is resumed to mimic a
restarted outburst.
\label{fig:w-t100-z}}

\end{document}